\documentclass[twocolumn, showpacs,preprintnumbers,amsmath,amssymb, longbibliography]{revtex4-1}
\usepackage{graphicx}
\usepackage{dcolumn}
\usepackage{bm}
\usepackage{natbib}

\begin{document}

\title{Intermittent dilation and its coupling to stress in discontinuous shear thickening suspensions} 
\author{Rijan Maharjan}
\author{Ethan O'Reilly}
\author{Thomas Postiglione}
\author{Nikita Klimenko}
\author{Eric Brown}
\email{ericmichealbrown@gmail.com}
\affiliation{Department of Mechanical Engineering and Materials Science, Yale University, New Haven, CT 06520}


\date{\today} 

\begin{abstract}
We investigate dilation-induced surface deformations in a Discontinuous Shear Thickening (DST) suspension to determine the relationship between dilation and stresses in DST.   Video is taken at two observation points on the surface of the suspension in a rheometer while shear and normal stresses are measured.   A roughened surface of the suspension is observed as particles poke through the liquid-air interface, corresponding to dilation.  Dilation events are found to be intermittent and localized spatially.  Shear and normal stresses also fluctuate between high- and low-stress states, and dilation is observed frequently in the high stress state.    On the other hand, a complete lack of dilation is observed when the stresses remain at low values for a several seconds.  Dilation is most prominent while the stresses grow from the low-stress state to the high-stress state, and the dilated region tends to span the entire surface by the end of the stress growth period.   Dilation is found only at stresses and shear rates in and above the shear thickening range.  These observed relations between surface dilation and stresses confirm that dilation and stresses are coupled in the high-stress state of DST.
\end{abstract}

\maketitle

\section{Introduction}
\label{sec:intro}

Discontinuous Shear Thickening (DST) is a transition in which the effective viscosity of a fluid increases sharply with shear rate in a steady state flow.  DST is generally found in suspensions of hard particles  without strong interparticle forces, and at packing fractions just below the liquid-solid transition (for a review, see \cite{BJ14}).  DST is often described as a transition from a low-stress state to a high-stress state, which occurs when shear forces overcome interparticle forces to cause contact and friction to form system-spanning contact networks \cite{BJ12, SMMD13, WC14}. The stress in the low-stress state can be dominated by hydrodynamic forces \cite{BB85, WC14}, weak repulsive forces \cite{Ho74,MW01a} attractive forces \cite{BFOZMBDJ10}, or other interparticle forces such as the weight of settled particles \cite{BJ12}.  The high-stress state is usually described as frictional, but there is debate over the relevance of dilation.  One model proposes that the high-stress state is a hybrid viscous-frictional bulk state where shear stresses are proportional to both normal stresses and shear rate \cite{WC14}, while another model proposes that the shear stress is only proportional to the normal stress, which is usually determined by the boundary  as it provides a restoring force in response to dilation (e.g.~from surface tension) \cite{BJ12}.  In this manuscript, we take video of the surface of a DST suspension during rheological measurements to search for any relationships between dilation and stresses.

It has long been claimed that DST in suspensions is related to Reynolds dilatancy, so much so that they are sometimes used as synonyms for each other \cite{MW58, Ba89}.  Reynolds dilatancy is an expansion in the volume of dense particle packing in response to shear.   As particles try to move past each other in a dense packing, they push against each other and the volume taken up by the particle packing tends to expand.  In a suspension, the liquid retreats from the surface to fill the larger voids that opened up between particles during dilation.  This is visible on the surface of the suspension as particles poke through the liquid-air interface to make the surface rough. This roughened surface reflects light diffusivity and appears matte when the particles are too small to see by the naked eye, but larger than around the wavelength of light. 

Dilatancy has been proposed to play an important role in shear thickening as the dilating material pushes against the boundary.  The boundary then responds with a restoring force, assuming the particles are harder than the least stiff boundary -- usually this is the liquid-air interface \cite{BJ12}.  As particles poke through the liquid-air interface, they deform the liquid-air interface with a minimum radius of curvature on the order of the particle size.  Surface tension provides a normal stress pushing back on the particles with a maximum on the order of surface tension divided by particle size \cite{CHH05, BJ12}.  This normal stress is transmitted from particle to particle along frictional contacts in system-spanning networks that repeatedly form and break up while the system is sheared.  The maximum stress supported in the shear thickening range has been shown to agree with this order-of-magnitude scaling for dozens of suspensions \cite{BJ12}.

Since dilation represents an expansion, dilation is often associated with normal stresses pushing against rheometer plates.  Normal stresses pushing against rheometer plates are reported in numerous experiments (for a summary, see \cite{BJ14}).   A few measurements of dilation based on volumetric change in DST suspensions exist as well.    Volumetric expansion at the liquid-air interfaced has been measured with a camera to be up to about 1 particle diameter, and was found to follow a predicted function of shear stress due to surface tension in response to dilation at the boundary \cite{BJ12}.  In a variation on the typical fixed-suspension-thickness rheometer experiment, controlled normal stress experiments which allowed the suspension thickness to vary measured a volumetric expansion with shear \cite{FHBOB08}.  While positive normal stress and corresponding dilation have been observed in DST suspensions, these observations do not distinguish between the models of a bulk rheology or a boundary-limited normal stress.
 
 
Published images of dilation at surfaces of DST suspensions are rare, and this may be part of why the dilatancy picture is not universally accepted.  Some of the early work discussing the relation between dilatancy and shear thickening is summarized by claims (but no images) that visible dilatancy at the surface of suspensions was observed starting at the same shear rate that shear thickening starts  \cite{MW58}.    One published  image of a suspension with ripples on the surface was associated with dilation \cite{OM00}, but appears to be a different surface instability than poking of particles through the liquid-air interface.  A close-up view of a suspension surface with a microscope showed particles poking through the liquid-air interface under shear in the high-stress state  \cite{BJ12}, but this was only shown at one applied stress.
Thus, while there are many claims, no evidence has been published which shows that the parameter range of shear thickening coincides with the parameter range of visible surface dilation.  

Perhaps the best images of what a dilated suspension surface looks like come from extensional flows.  Smith showed images of a  matte (i.e.~rough) surface for suspensions under extension for a high strain rate, contrasting with a lower strain rate where no such surface change was observed \cite{SBCB10}.   A follow-up to these extensional experiments reported that dilation was observed after the force started to increase, but before peak force was reached \cite{Smith15}. 
 We note that many of these  images show shear jamming (a.k.a~granulation) after flow has stopped rather than dilation during a steady shear, and while shear jamming is often found to occur in DST suspensions, it is not equivalent to DST \cite{JHLJJ18}.  Impact experiments have also shown roughened surfaces as a result of dilation  \cite{RMJKS13,  ASMMB18}.  It is perhaps suspicious that the best images of surface dilation are from extension and impact  experiments -- even though there are some similarities between shear thickening in steady state shear and transient extensional flows, there are also significant differences,  for example transient extensional flows can result in orders-of-magnitude larger stresses than steady state shear \cite{MMASB18, MAB18}.  It remains an open question whether dilation at the surface in extensional flows applies straightforwardly to shear flows.   

While shear thickening is described by a steady state mean viscosity curve, time-resolved measurements reveal that both shear and normal stresses fluctuate, sometimes by orders of magnitude \cite{LDH03, LDHH05}.  Despite the strong fluctuations, there remains a proportionality between shear and normal stresses with a coefficient of order 1, suggestive of a Coulomb frictional relationship between stresses \cite{LDH03, LDHH05, BJ12}.  Shear thickening in the mean viscosity curve can be attributed to the contribution of fluctuations to the high-stress state, while a low-stress state follows the trend of a viscosity curve without shear thickening  \cite{LDH03}.  Stress fluctuations have also been found to occur in localized regions, rather than be uniform throughout the suspension \cite{RBU17}.   If dilation at the suspension surface is related to stresses \cite{BJ12}, then the fluctuations and localization of stresses should have a connection to dilation, but no such fluctuations or localization in surface dilation have ever been reported in steady-state flows.

In this work, we present images and video of the surface of DST suspensions along with simultaneous rheology measurements under steady shear.  We focus on images at the surface because it is the deformation of the liquid-air interface that is predicted to more directly affect the rheology \cite{BJ12}, as opposed to changes in the bulk packing fraction.  Specific issues we address are (1)  to present clear video and images of dilation under steady shear, (2) show whether surface dilation occurs in the same stress range as DST, and (3) determine whether surface dilation is localized and fluctuates along with stresses.  

The remainder of the manuscript is organized as follows.  Section \ref{sec:methods} describes the apparatus, materials, and rheometer measurement procedure. Section \ref{sec:observations} shows images of dilation at the suspension surface and describes observations of intermittency and localization.  Section \ref{sec:dilationmethods} explains image analysis methods to quantitatively track dilation over time from the video.  Section \ref{sec:timeseries} shows simultaneous time series of stress and dilation measurements to identify their relationship during fluctuations.  Section \ref{sec:statistics} presents statistics on the probability of dilation under different conditions such as at high- and low-stress states.  Sections \ref{sec:visccurve} and \ref{sec:dilation_stress} present measurements over a range of shear stress and stresses to confirm that dilation occurs at stresses and shear rates in the shear thickening range and above.  Section \ref{sec:discussion} discusses an interpretation of time series in analogy to  stress growth and failure in material stress-strain curves, and some of the consequences of a rheology dependent on boundary stiffness as opposed to bulk properties.

\section{Materials and methods}
\label{sec:methods}

\subsection{Materials}

We used cornstarch and water as an example of a typical DST fluid \cite{BJ14}.  Cornstarch was purchased from Argo, and suspended in tap water near room temperature. The suspensions were mixed and measured at a room temperature of $22.0\pm0.6$ $^{\circ}$C and humidity of $48\pm6\%$, where the uncertainties represent day-to-day variations in the respective values.  A four-point scale was used to measure quantities of cornstarch and water to obtain a weight fraction $\phi_{wt}$.  Each suspension was mixed until no dry powder was observed. The suspension was further shaken in a Vortex Genie 2 for approximately 30 seconds on approximately 60\% of its maximum power output.

Packing fractions of cornstarch suspensions are difficult to compare, mainly due to the different adsorption of water into nominally dry cornstarch at different temperature and humidity \cite{MB19}.   For comparison to data from other researchers in the DST range, we use an effective packing fraction scale $\phi_{eff}$ that is based on the measured shear rate $\dot\gamma_c$ at the onset of DST, which is more precise than comparing values based on measured weight fractions in different environmental conditions for packing fractions in the DST range, due to the divergence of properties near the liquid-solid transition at $\phi_c$ \cite{MB19}.   We use the effective packing fraction $\phi_{eff}/\phi_c = 1-0.0475\dot\gamma_c^{0.268}/d$ in the range 0.61 mm $\le d \le 1.8 $ mm, calculated from the suspension thickness $d$, measurements of the shear rate $\dot\gamma_c$ at the onset of DST, and measurements of $\phi_c$ under current environmental conditions \cite{MB19}.

\subsection{Apparatus}

\begin{figure}
\centering
\includegraphics[width=0.475\textwidth]{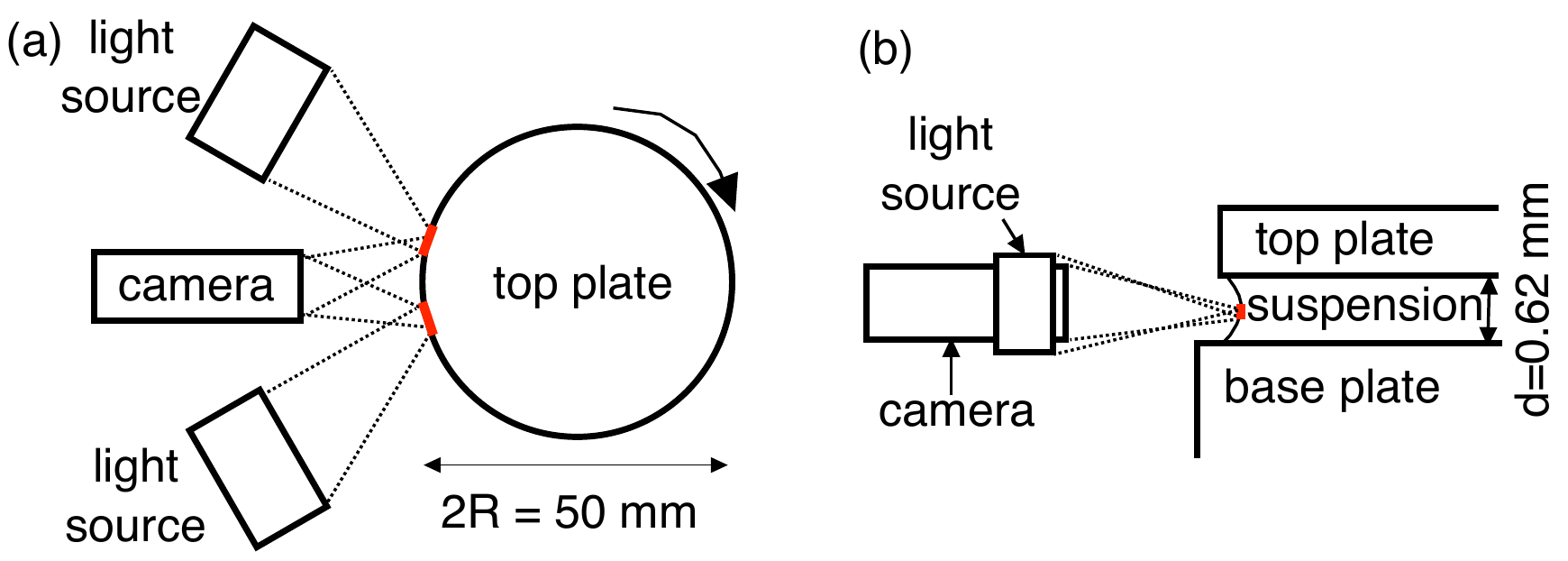}
\caption{(color online) Illustration of the experimental setup (not to scale). (a) Top view.    (b) Side view.  The suspension is sheared underneath the rheometer top plate.  Dotted lines illustrate paths of light rays from the edges of the light source to the edges of the camera sensor.  The camera views direct reflections  as bright spots (red) on the suspension surface.  
When the suspension dilates, the roughened surface produces a diffuse reflection over a wider area of the suspension surface.
}
\label{fig:setup}
\end{figure}

Experiments involved a combination of rheological measurements with high speed video of the suspension surface, as illustrated in Fig.~\ref{fig:setup}.  Rheological measurements were performed with an Anton Paar MCR 302 rheometer with a suspension placed between parallel plates.  The rheometer measured the torque $M$ on the top plate  and  angular rotation rate $\omega$  of the top plate.   In  different experiments, either torque or rotation rate could be controlled,  while the other was measured as a response.   The spatially averaged shear stress is given by $\tau = 2M/\pi R^3$ where $R$ is the radius of the top plate.  While the mean shear rate varies along the radius of the suspension, the mean shear rate at the edge of the plate is used as a reference parameter, which is given by $\dot{\gamma} = R\omega/d$ where $d$ is the suspension thickness between the top and base plates.  The viscosity of the suspension is measured as $\eta = \tau/\dot{\gamma}$ in a steady state.   We also measured the force $F$ on the top plate of the rheometer (upward positive) and report the mean normal stress $\tau_N = F/(\pi R^2)$ with a resolution of 2.5 N.   To calibrate the normal stress, we subtract the normal stress on the suspension at rest, which is due to surface tension at the liquid-air interface.   

Since our measurements involved imaging of the suspension-air interface, special consideration was given to how the placement of the suspension affects the suspension-air interface.  In steady shear experiments, there is a tendency for the side surface of the suspension to become unstable, in particular, blobs of suspension build up on the side and protrude outward temporarily, interfering with imaging.  We found this effect was minimized if the suspension did not extend all the way out to the edge of the plate, as shown in Fig.~\ref{fig:setup}b,  which presumably confines the suspension in place better by providing more surface area on the parallel plate surfaces for the contact line of the liquid-air interface.  Therefore, when loading the suspension we adjusted the top plate height so the suspension had a mean radius of $24.75\pm0.25$ mm compared to the top plate radius of $R=25.00$ mm.  This results in up to a 6\% underestimate in shear stress $\tau$ and up to a 4\% underestimate in normal stress when the plate radius $R$ is used in calculations.   The resulting range of suspension thicknesses was typically $d=0.62\pm 0.02$ mm, where the $\pm$ corresponds to the standard deviation of different experiments, unless otherwise noted.  

The camera and lights were arranged as shown in Fig.~\ref{fig:setup} to visualize the side of the suspension.  High resolution images, videos, and examples shown in Sec.~\ref{sec:observations}, \ref{sec:dilationmethods}, and \ref{sec:dilation_stress} were taken with a DSLR camera with a micro lens.   A small amount ($<0.1$\%) of powdered graphite was added to suspensions to visualize shear for high resolution images and video. High time resolution measurements for comparing the timing of stress fluctuations with dilation reported in Sec.~\ref{sec:timeseries}-\ref{sec:visccurve} were taken with a Phantom Miro M110 camera with a micro lens.
To obtain the highest resolution image possible, we placed the camera as close to the suspension surface as possible while remaining in focus.   One or two light sources were placed at angles such that direct reflection from the suspension surface would be observed at the camera.  Such direct reflection was found to produce the highest contrast between the shiny surface at rest and a rough surface with a diffuse reflection when the suspension dilates.   The angles of the light sources were also set to produce relatively large bright spots to increase the fraction of surface observed.  For quantitative measurements in Sec.~\ref{sec:statistics} these bright spots were 5 mm long and separated by 7.5 mm.  The resolution was 20 pixels across the suspension thickness $d$, and 640 pixels in the horizontal direction.  This viewed about 12\% of the suspension surface, although only about 6\% of the circumference of the surface was reflecting bright spots.  Video data was recorded as 8-bit grayscale images. 

A solvent trap (not shown) was placed around the suspension to minimize moisture exchange between the suspension and the atmosphere.  The solvent trap effectively placed a water seal around the  suspension, with a lipped lid touching water contained on the top, cupped, surface of the top plate.  This was specially designed with a square horizontal cross-section made of transparent acrylic to avoid distortion of the images from light passing through the walls of the solvent trap. 


The lights used for taking video heated up the suspension during measurements, and the custom solvent trap required a base that interfered with the rheometer temperature control, resulting in suspension temperatures in the range $23.5\pm1.5^{\circ}$ C for different experiments, measured at the rheometer plate.  

\subsection{Measurement procedure}

A preshear was used to eliminate effects of loading history and acceleration on the suspension to produce steady-state measurements.  The preshear consisted of a linear ramp in shear rate or stress from rest to a value above the shear thickening range, lasting at least 200 s.  Steady-state measurements started immediately after this,  at the same shear rate or stress the preshear ended at to minimize acceleration.

The first and last measurement on each suspension was a viscosity curve, which was used to identify the shear thickening range.   The critical shear rate $\dot\gamma_c$ and shear stress $\tau_c$ are defined to be at the onset of DST, obtained at the lowest shear rate $\dot\gamma$ where $\partial\eta/\partial\dot\gamma>1$ \cite{MB19}.   The shear rate $\dot\gamma_{max}$ and shear stress $\tau_{max}$ are defined as their maximum values in the shear thickening range, i.e.~at the local maximum of viscosity. 

Viscosity curves were measured with a controlled shear rate, and the ramp was done at a rate of 300 to 1000 s per decade.  Each ramp was immediately remeasured in reverse order to confirm there was no hysteresis or time-dependent effects.  With the last measurement being a repeat of the viscosity curves, we checked the magnitude of any time-dependent effects such as evaporation over the course of measurements.  Over the course of our  longest series of experiments of 4 hours, the measured value of $\dot\gamma_c$ decreased by 30\%, corresponding to an increase of up to 0.008 in $\phi_{eff}/\phi_c$ in the DST range.




The relative timing of the video and stress measurements was done manually by triggering the video once the steady state is reached in rheometer measurements.  This is referred to as time $t=0$ in both measurements.  This process results in an uncertainty of about 1 s in the relative time between the two measurements reported in Secs.~\ref{sec:timeseries} and \ref{sec:statistics}.

\section{Observations of dilation}

\label{sec:observations}

\begin{figure*}
\centering
\includegraphics[width=1\textwidth]{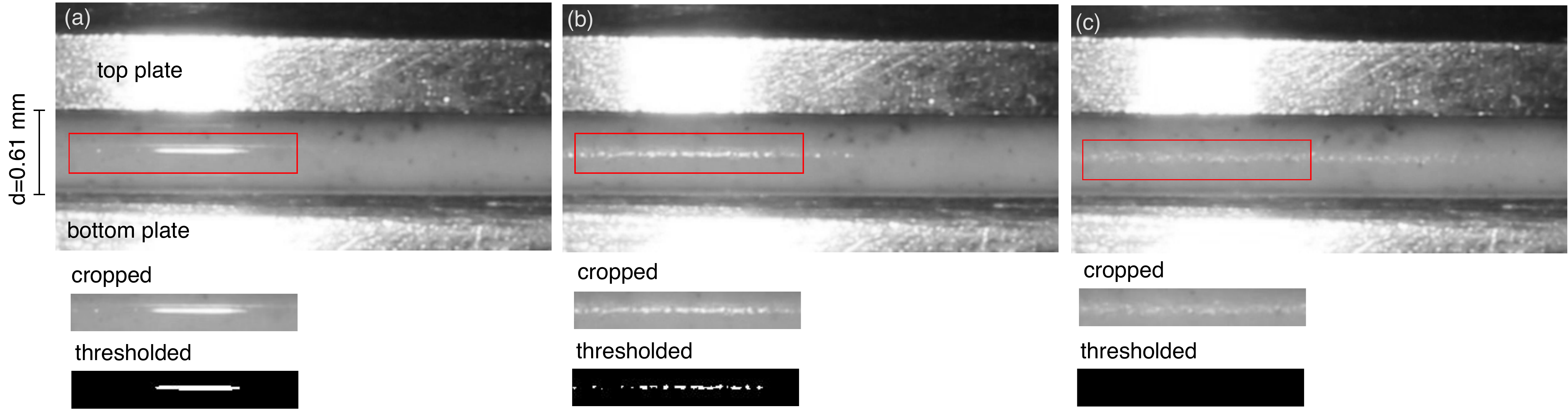}
\caption{(color online) Side view of the suspension in the rheometer. (a) A suspension at rest with no surface deformation, indicated by the shiny surface. (b)  Surface of a suspension under shear as the dilated region is growing.  Dilation is indicated by the speckled pattern, due to the rough surface reflecting light diffusively. (c) Surface of a fully dilated suspension under shear.  Lower images illustrate the cropping and thresholding used to identify presence of dilation based on counts of the number of bright pixels. The red boxes in the top images outline the cropped region.}
\label{fig:raw_images}
\end{figure*}

An example video of the suspension surface is shown in Supplementary Video 1  for an applied shear stress $\tau=200$ Pa and effective packing fraction $\phi_{eff}/\phi_c=0.95$, in the middle of the DST range, played back at 20x real speed.  Snapshots of different behaviors are also shown in Fig.~\ref{fig:raw_images}.  We focus our attention on the reflection of light from the surface of the suspension, as that is closely related to deformation of the liquid-air interface, which  is predicted to be the relevant consequence of dilation.  For a suspension at rest, and much of the time during steady shear, there is a thin bright strip of light reflected off the surface, as illustrated in Fig.~\ref{fig:raw_images}a.  This is the direct reflection from the light source off of a smooth, shiny surface. The bright spot is a thin strip even though the light source is circular because the resting liquid-air interface is highly curved in the vertical direction due to surface tension, and less curved in the horizontal direction due to the plate curvature, as illustrated in Fig.~\ref{fig:setup}.     During shear, the bright spot can fluctuate in size, shape, and position somewhat due to slight deformations of the surface.  These fluctuations may be due to vibrations of the experimental apparatus


The effects of dilation on a suspension during shear can be observed in Fig.~\ref{fig:raw_images}b and Supplementary Video 1.  Instead of a single bright spot, the surface now appears speckled over a wider area.  The  change in the reflection of light indicates a deformation of the surface of the suspension.  More specifically, this can be understood as a consequence of dilation according to the following argument.  When a dense suspension is sheared and dilation occurs, the particle packing expands (known as Reynolds dilatancy), resulting in particles poking through the liquid-air interface, while liquid retreats into the voids opened up in the interior.  Since the average particle diameter of 14 $\mu$m  is too small to see individual particles, but larger than the wavelength of light, the direct lighting reflects diffusively off the roughened surface, resulting in the speckling seen in Fig.~\ref{fig:raw_images}b and c.  The surface appears progressively rougher and dimmer because there is less direct reflection. The region where reflected light is coming from also spreads out horizontally and vertically because the roughened surface results in additional localized direct reflections from the light source to the camera where there were none before.  A larger, dimmer speckled region indicates a larger surface deformation to reflect light at wider camera angles.

\subsection{Intermittency}


Supplementary Video 1 shows intermittent changes in the surface reflection. Specifically, we occasionally observe dilation events where the images transition from the bright spot seen in Fig.~\ref{fig:raw_images}a into the speckled dilation pattern shown in Fig.~\ref{fig:raw_images}b on the way to the dimmer speckled pattern in Fig.~\ref{fig:raw_images}c, before brightening again and reverting back to a undeformed surface.  

\subsection{Localization and Propagation}

\begin{figure}
\centering
\includegraphics[width=0.475\textwidth]{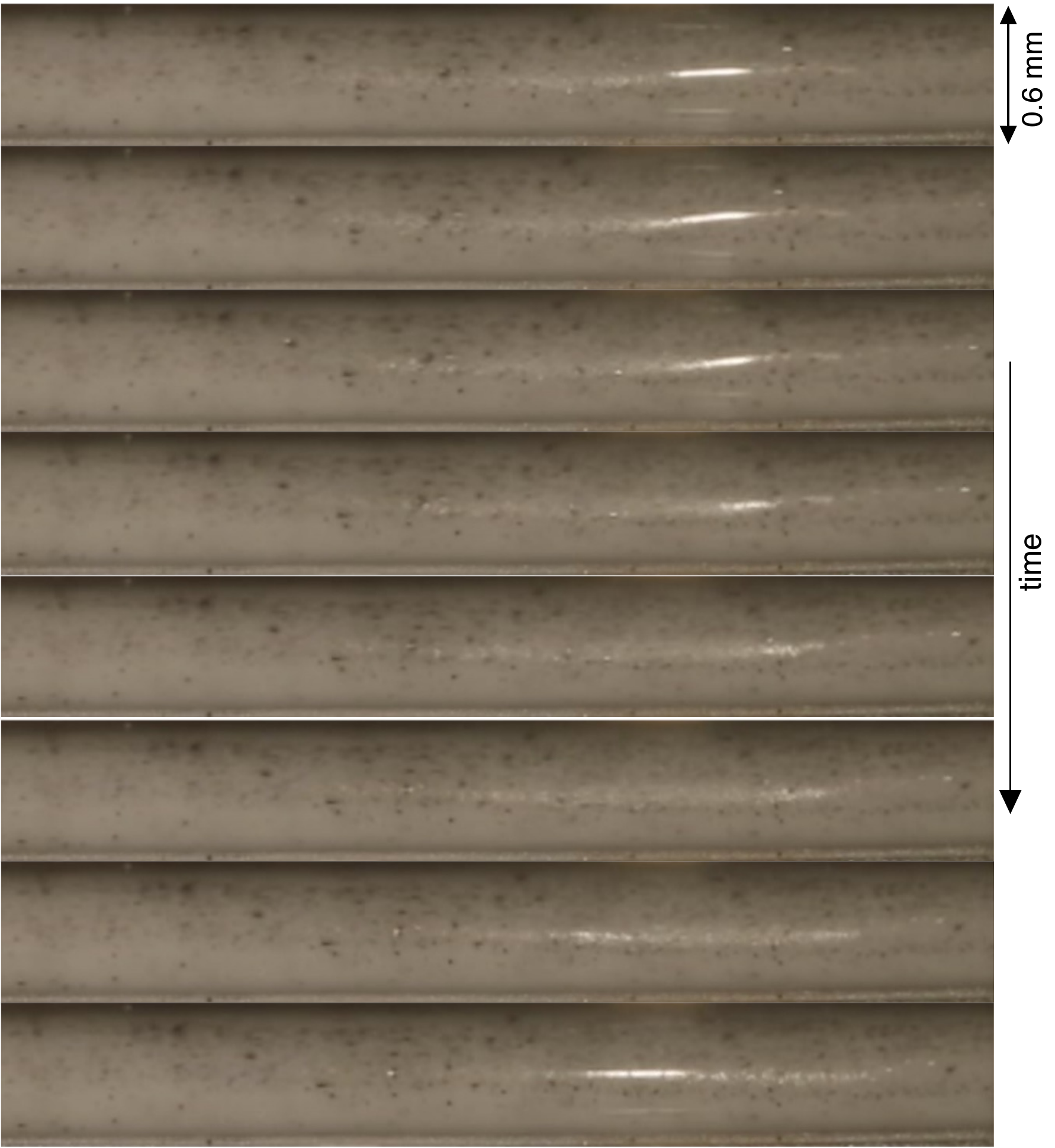}
\caption{A series of images of the suspension at different times separated by 0.033 s, showing the propagation of a localized dilated region (the faint speckled pattern) past the bright spot from left to right.    
}
\label{fig:frontpropagation}
\end{figure}

 Supplementary video 2 shows examples of dilated regions that are localized  i.e.~they don't have to span the entire system at once, and they can grow or propagate over time.  This data was taken at a constant shear stress of $\tau=100$ Pa and $\phi_{eff}/\phi_c=0.90$, in the low end of the DST range where these localized events are more apparent, and the video is played back at 4x real speed.  A series of images in  Fig.~\ref{fig:frontpropagation} show an example of localized dilated region propagating past a bright spot, although the series of images is not nearly as clear as the video. The series of images shows a dilation event first appearing to the left of the bright spot, and propagating to the right.  Such propagation events are numerous, but are more frequent at higher stress.   From the video it can be seen that these fronts propagate at a constant speed, and faster than the plate speed.  Dilated regions propagate in both directions, although the majority propagate in the direction opposite the top plate rotation.  This contrasts with measurements of localized high stress regions found to propagate through the interior in stress-controlled measurements \cite{RBU17}, where the intermittent and localized stress fluctuations propagated at 1/2 the plate speed and in the direction of plate motion, due to advection.


Although not all dilated regions are the same size, it appears that they all have a characteristic size at a given packing fraction and shear rate.  Our preliminary investigations into lower packing fractions $\phi_{wt}/\phi_c \le 0.89$ in the Continuous Shear Thickening (CST) range (defined by viscosities that increase with shear rate, but less steeply than DST \cite{MB19}) resulted in smaller dilated regions, often smaller than the bright spot in our images.  However, the instability of those suspensions at high stress limited us to measuring at lower stress  and prevented us from collecting systematic packing-fraction-dependent statistics to compare to the higher stresses reported here.  
 With small dilated regions and 2 bright spots, we occasionally observe 2 separate dilation events at the same time  (a few percent of events), so there can be more than one dilation event occurring at once, and there may be even more on other portions of the surface that are not in our viewing range.   In contrast, the dilation events in Supplementary Video 1 do not have  a clear propagation direction, but rather appear to span our measurement region.  This may correspond to a limit where dilated regions are significantly larger than the measurement region, and with growth rates that are faster than we can resolve, perhaps due to the higher packing fraction of that data.



\section{Methods to quantitatively track dilation in videos}
\label{sec:dilationmethods}

\subsubsection{Obtaining the number of bright pixels $n$}

We identify a quantitative measure of dilation from images so that we can calculate statistics and compare time series to those of stresses later.    Because the bright spot is a result of direct reflection off a nearly-flat surface, we found that the absence of bright pixels is a good proxy for surface dilation.  The lower panels of Fig.~\ref{fig:raw_images} illustrate the process of our image analysis.  We first crop images as shown in Fig.~\ref{fig:raw_images} to remove the rheometer plates from the image, while remaining large enough to contain the bright spot from the direct reflection from the light source, as well as the region it moves around in during experiments.  In some experiments, where dilated regions tended to be smaller than the bright spot, we cropped images down to a size smaller than the dilated region so that dilation events would consistently result in nearly zero bright pixels. 
We then convert the cropped grayscale images into binary black and white images.  We set a brightness threshold such that pixels with greater or equal  brightness values are converted to white, and pixels with lower brightness values are converted to black, as shown in the lower panels of Fig.~\ref{fig:raw_images}, chosen to ensure that the bright reflected spot always maintains many white pixels, while fully dilated regions are converted to black pixels.   Once we have cropped and thresholded the image, we count the number of bright pixels $n$.  We note that statistics of $n$ depend on the optical and lighting setup and threshold value, so such statistics can only be meaningfully compared to other datasets with similar methods. 

\subsubsection{Time series of the number of bright pixels $n$}

\begin{figure}
\centering
\includegraphics[width=0.475\textwidth]{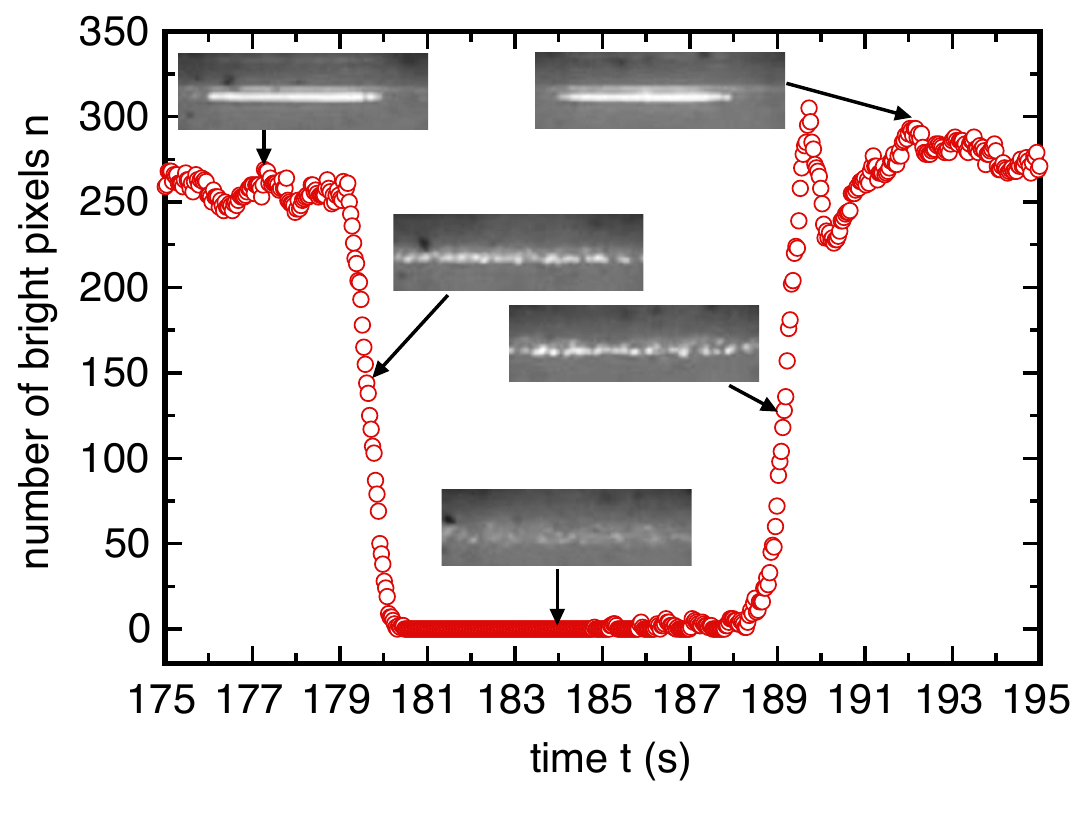}
\caption{Number of bright pixels $n$ as a function of time for a single dilation event, showing the corresponding raw images in the experiment. The low values of $n$ correspond to when surface dilation is observed. 
}
\label{fig:dilation_stages}
\end{figure}

We show a short example time series of the number of bright pixels $n$ for a typical dilation event in Fig.~\ref{fig:dilation_stages} for a controlled shear stress $\tau=200$ Pa  (average shear rate $\dot\gamma=0.12/$s) at $\phi_{eff}/\phi_c=0.945$, in the DST range.   Cropped snapshots of the surface are shown at a few points in time.   Initially, when $n$ is high, the snapshot shows a single bright spot, indicating a smooth surface, similar to Fig.~\ref{fig:raw_images}a.  There is a distinct dilation event where the number of bright pixels drops rapidly to near zero for a while, before returning to its high value.   When $n$ is at intermediate values, the snapshots show speckled patterns due to a rough surface, similar to Fig.~\ref{fig:raw_images}b.   When $n=0$, the snapshot shows only dim, diffuse reflection corresponding to a completely dilated state,  similar to Fig.~\ref{fig:raw_images}c.   We manually checked that each dilation event that we could see in a video resulted in a small $n$, while $n$ remained large when dilation was not seen in videos.  This confirms that dilation events can be distinctly tracked based on the number of bright pixels as a function of time.

\begin{figure}
\centering
\includegraphics[width=0.475\textwidth]{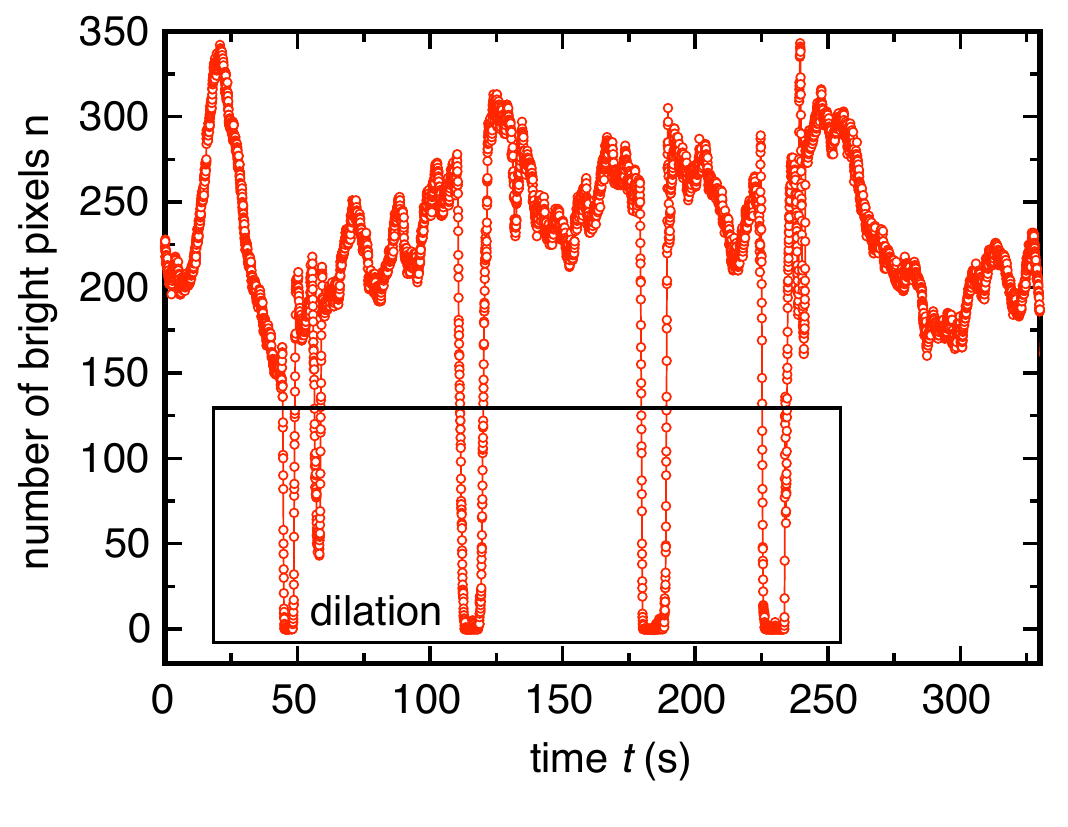}
\caption{Number of bright pixels as a function of time showing several dilation events (indicated by the box), when $n$ drops to near zero. Dilation events are found to be intermittent.
}
\label{fig:raw_pixels}
\end{figure}

A longer time series from the same dataset is shown in Fig.~\ref{fig:raw_pixels}.  We observe that when the suspension is not dilated, $n$ fluctuates in a preferred range of $\approx 200-300$.  The fluctuation in $n$ is due to slight surface deformations that cause the size of the bright spot to vary.  These slight fluctuations in bright spot size are visually distinct from the rough, dilated surface.  Figure \ref{fig:raw_pixels} shows five distinct dilation events where $n$ drops to near zero.   These dilation events occur intermittently, and while we sometimes observe a preferred time between events, the time series are not strictly periodic.  There also appears to be a typical slope $\partial n/\partial t$ during the transitions, indicating a consistent growth rate of dilated regions.  Each of these typical values varies with different applied stress $\tau$ or packing fraction $\phi_{eff}$, although we did not systematically investigate these parameter dependences here.

\subsubsection{Definition the threshold $n_t$ for dilation}
\label{sec:dilation_prob}

\begin{figure}
\centering
\includegraphics[width=0.475\textwidth]{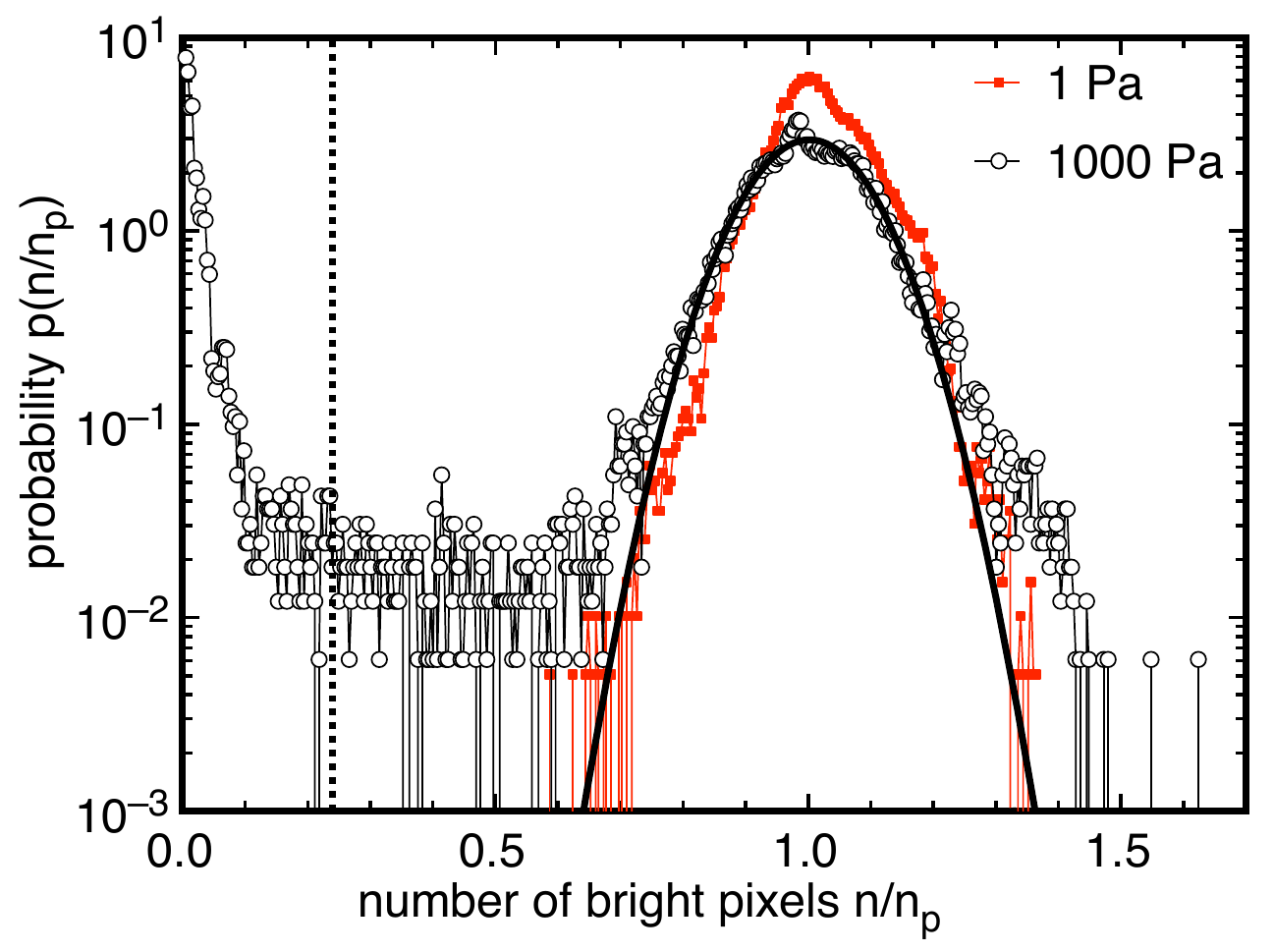}
\caption{Probability distribution of the number of bright pixels $n$.  Different applied shear stresses $\tau$ are indicated in the legend.  The broad peak at $n=n_p$ corresponds to the non-dilating state.  Solid line: fit used to determine the peak location $n_p$.  The sharp peak at $n=0$ corresponds to full dilation.    Data with no visible dilation at $\tau=1$ Pa does not have the peak at $n=0$ or the low-$p(n)$ plateau.  Dotted line: standard threshold value $n_t/n_p$ below which dilation is counted. 
}
\label{fig:probnumpixels_stress}
\end{figure}

To obtain statistics on dilation, we define a threshold value of $n$ below which dilation is said to occur.  To identify this threshold, we use features of the probability distribution of $n$.  A probability distribution of $p(n/n_p)$ is shown in Fig.~\ref{fig:probnumpixels_stress} from an extended time series at $\tau=1000$ Pa and $\phi_{eff}/\phi_c = 0.905\pm0.008$, at the low end of the DST range.  
 $n$ is normalized by its value at the non-zero local maximum $n_p$ corresponding to the non-dilated state, determined by a Gaussian fit as shown in Fig.~\ref{fig:probnumpixels_stress}.  This normalization provides a scale in which experiments under different lighting conditions could potentially be compared.  The peak at $n/n_p=1$ is wide due to the fluctuations in the size of the bright spot due to slight surface deformations.  The sharp peak at $n=0$ corresponds to dilation events with no bright pixels.   There is a low plateau in $p(n)$  for $0.10 \stackrel{<}{_\sim} n/n_p \stackrel{<}{_\sim} 0.68$.  This corresponds to a range that is only traversed rarely as the system is transitioning during dilation events from a big bright spot to fully dilated, as seen in Figs.~\ref{fig:dilation_stages} and \ref{fig:raw_pixels}.  We also confirm that datasets with no visible surface dilation have no peak in $p(n)$ at $n=0$ or low-$p(n)$ plateau, as shown for $\tau =1$ Pa in Fig.~\ref{fig:probnumpixels_stress}.


 For purposes of obtaining statistics on dilation, we define dilation as occurring when $n$ is less than a threshold value $n_t=0.24n_p$ for all data presented in this paper.  
 This threshold value is in the low-$p(n)$ plateau for all datasets where we observe dilation events, even for data taken with a different camera or optical setup.
 While the choice of the value of $n_t$ is somewhat arbitrary within the broad low-$p(n)$ plateau, the low probability in this plateau means that the probability of dilation is least sensitive to the threshold value $n_t$.   Again, because of the somewhat arbitrary nature of this threshold value, statistics can only be meaningfully compared to other datasets with the same optical and lighting setups.   Distinct dilation events are identified as a continuous segment of the time series with $n<n_t$. 


\section{Correlation between shear rate fluctuations and observations of dilation}
\label{sec:correlation}

\subsection{Time series of stress measurements and surface reflections}
\label{sec:timeseries}

\begin{figure*}
\centering
\includegraphics[width=1\textwidth]{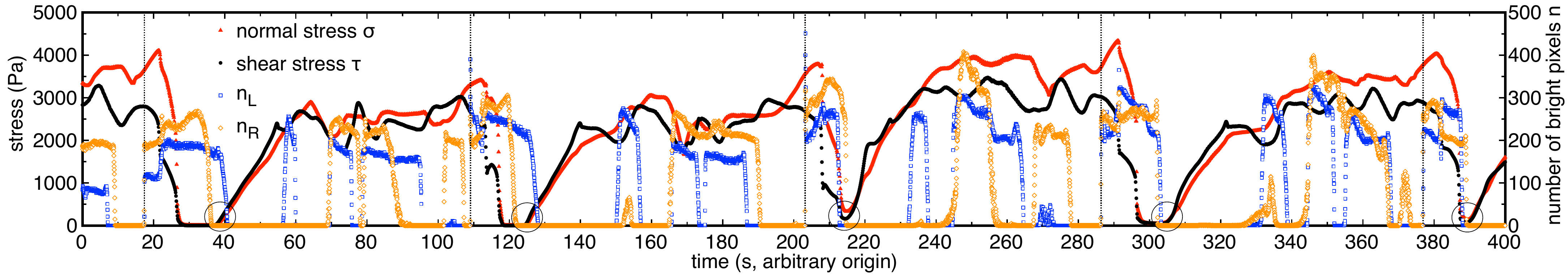}
\caption{(color online) A time series showing the relation between dilation and stress. Left axis: shear stress $\tau$ (solid circles) and normal stress $\sigma$ (solid triangles).   Right axis:  number of bright pixels in each bright spot, $n_L$ (open squares) and $n_R$ (open diamonds).  The start of stress growth events are indicated by large open circles.  Both $n_L$ and $n_R$ drop to zero within a few seconds of each of these events, indicating the stress increases occur along with dilation.  Vertical dashed lines:  Times where both $n_L$ and $n_R$ increase simultaneous from zero and shear stress begins to drop, indicating that stress drops occur when dilation ends.
 }
\label{fig:timeseries}
\end{figure*}

The relation between dilation and stress fluctuations is shown in a time series of the number of bright pixels along with simultaneous stress measurements.   Figure \ref{fig:timeseries} shows data for $\phi_{eff}/\phi_c=0.952$, in the DST range, and at a controlled shear  rate $\dot\gamma=0.2$ s$^{-1}$ $\approx 1.2 \dot\gamma_{max}$, so that the mean stress corresponds to the high-stress state of DST.  For these statistics, we use two light sources to generate two bright spots.  Bright spot analysis is done separately for each observation location, and we report time series for the number of bright pixels $n_L$ for the left spot, and $n_R$ for the right spot.
Several stress drop and increase events can be observed as the stresses fluctuate between the low and high-stress states of DST.  For each stress drop and increase, the normal and shear stress tend to track each other fairly well.  Each time the stresses starts to drop to near zero, there is also a sharp increase in both $n_L$ and $n_R$ from zero up to near the non-dilated state value $n_p$ (indicated by the vertical dashed lines in Fig.~\ref{fig:timeseries}).  This indicates that the stress drops tend to start when dilation ends.  On the other hand, each time the stresses start to increase from near zero toward their mean values, both $n_L$ and $n_R$ drop to zero within a few seconds of the start of the stress increase (these times are marked by the large open circles in Fig.~\ref{fig:timeseries}).  This indicates that stress increases tend to occur along with dilation.  Together, these observations indicate a strong correlation between dilation and the high-stress state of DST suspensions.

While we tend to see dilation occur along with high stresses  in Fig.~\ref{fig:timeseries}, there are also many instances where dilation disappears at the observation locations while the stresses remain high.  
Since we only observe a small fraction of the surface, we likely miss a number of dilation events, which may exist in the remaining portions of the high stress periods.  In contrast, while the stress remains low, we observe a complete lack of dilation events, except when they are within a few seconds of a stress increase.  These observations are at least consistent with the hypothesis that the high-stress states require dilation somewhere on the surface.

The drops in $n_L$ and $n_R$ do not coincide as well with each other near the time of stress increase as they do near the time of stress drops, suggesting that when a stress drop occurs, dilated regions shrink rapidly, while during stress increases, a longer time is required for surface dilation to propagate around the suspension.    The growths in stress to the high-stress state also tend to be slower than the drop in stress to the low-stress state.  These observations indicate an asymmetry in the rates of buildup and breakdown of high-stress-state structures.




\subsection{Statistics relating dilation and stress}
\label{sec:statistics}

To collect more statistics on dilation events, we measured stresses and surface dilation over time on six different suspensions.  Three of the suspensions measured correspond to packing fractions in the DST range ($\phi_{eff}/\phi_c=0.952, 0.953, 0.957$), with applied shear rates $\dot\gamma \approx 1.2\dot\gamma_{max}$, where the mean stress is close to the high-stress state.   We calculated the probabilities of dilation occurring at each bright spot $p_1$ under different conditions to quantify the likelihood of dilation in different states such as the high-stress or low-stress states. The following numbers in this subsection are averaged (weighted by timestep) over the three datasets in the DST range where we observed a high-stress state.  Probabilities of dilation $p_1$ are summarized in Table \ref{tab:dilation_prob}.

\begin{table}
\begin{tabular}{lrrr}  
 \hline
 \hline
   & $p_1$ & $p_2$ & length $\lambda$ \\ \hline 
high stress ($>1400$ Pa) & $52\%$ & $40\%$ & 46 mm\\ \hline
stress growth  & $90\%$ & $86\%$ & 240 mm\\ 
stress growth for previous 9 s & $100\%$ & $100\%$ & $\infty$\\ \hline
stress drop & $7\%$ & $3\%$ & 9 mm\\ \hline
low stress ($<460$ Pa) & $21\%$ & $11\%$ & 15 mm\\ 
low stress, low growth & $12\%$ & $6\%$ & 14 mm\\ 
low stress, low growth for next 9 s & $0\%$ & $0\%$ & NA\\ 
CST (low stress) & 4\% & 0.08\% & -1 mm\\ \hline \hline
\end{tabular}
\caption{Probabilities of dilation at single-point measurements $p_1$, and simultaneously at two-point measurements $p_2$, compared for different shear stress conditions as explained in the text.   Dilation is common at the high-stress state, and uncommon in the low-stress state.  No dilation is found at times when both the stress and stress growth remain low for the following 9 s.  Dilation is most common during stress growth periods that transition from the low-stress state to the high-stress state, and is always found after consistent stress growth for 9 s.  The correlation length $\lambda$ of dilated regions is extrapolated from these probabilities.  The extrapolation suggests that the size of dilated region during the stress growth period typically spans the surface of circumference 157 mm.
 }
\label{tab:dilation_prob}
\end{table}

 In the high-stress state $p_1= 52\%$, where the high-stress state is defined as when the shear stress is above a threshold of 1400 Pa, just outside of the range of fluctuations observed in the high-stress state in Fig.~\ref{fig:timeseries}.   In contrast, in the low stress  state $p_1=21\%$, defined initially as when the shear stress is below the threshold of 460 Pa (25\% of the mean stress).  While the stress usually drops much lower, as low as the stress $\tau_c=2\pm1$ Pa at the onset of DST  \cite{LDH03}, it does not always reach that low level, so the stress thresholds for counting events are higher to include all significant stress drop events in our data.   These statistics indicate that dilation is significantly more likely in the high-stress state than in the low-stress state.

We refine the above statistics by identifying distinct transitions between the high and low-stress states and further subdividing the time series to include stress drop and stress growth periods.  Each transition to the low-stress state is defined to occur when the shear stress drops below 460 Pa, and the transition back to the high-stress state is defined when the shear stress next exceeds 1400 Pa.   Successive transitions to the same stress state are not counted until a transition to the other stress state, and the thresholds are set at different values to avoid counting extra events due to jitter of the time signal around the thresholds.   With this counting, we find a total of 50 such pairs of stress transition events and 537 dilation events among the three time series that reach the high-stress state.  

 We next consider statistics of dilation during the stress growth period corresponding to the transition from the low-stress state to the high-stress state, where the stress growth period is defined as when the shear stress is in between the thresholds of 460 Pa and 1400 Pa, and the five-point slope of shear stress over time is greater than 50 Pa/s (the mean slope seen in the stress growth regions of Fig.~\ref{fig:timeseries} is found to be 150 Pa/s for each dataset).   During this stress growth period $p_1=90\%$.   These probabilities of dilation are even higher than those in the high-stress state, indicating that dilation is correlated more strongly with stress growth than the high-stress state.  

The probability of dilation in the growth period increases to $p_1=100\%$ when we only include data where the stress growth exceeded 50 Pa/s for the preceding 9 s (this limited range of data still includes 1800 data points from 23 distinct stress growth periods).  The stress growth period required for $p_1=100\%$ may correspond to the maximum time it takes for a dilated region to grow and span the entire surface to ensure that we observe dilation at both measuring locations.  Since the average time required to grow up to the mean stress of the high-stress state of  2460 Pa is 16 s, this means that surface-spanning dilation would be consistently reached by the time the high-stress state is reached by sustained typical growth from the low-stress state.


Given the high correlation observed between dilation and stress growth, we narrow the definition of the low-stress state to times when the shear stress is below the threshold of 460 Pa, and the slope of stress over time is less than 50 Pa/s.    The probability of dilation in this low stress, low stress growth period is $p_1=12\%$.  This is  more distinct from the high-stress state, and suggests the dilation observed at low stress is probably more associated with the growth in stress over time. 

 If we further limit our analysis of statistics in the low-stress state to data points that satisfy the low stress and low stress growth condition for the following 9 s, then we find no dilation  ($p_1=0$).    This smaller range of data with no dilation still includes 8000 data points.  While 9 s is the longest persistence time required, the persistence time at low stress is usually much lower, for example, for a persistence time of 2 s,  $p_1$ drops to half of its value with no persistence time.  Since the dilation occurs at the end of low stress periods followed by stress growth, this time corresponds to the longest time it takes for the stress growth at the plates to respond to dilation.  Since this is similar to  the time it takes for dilation to span the surface after the stress starts increasing, it suggests this time scale of 9 seconds is the longest time it takes for a signal connecting surface dilation and shear stress to travel across the system, at least for the stress growth period.


We finally consider statistics of dilation during the stress drop period that transitions from the high-stress state to the low-stress state.  We find dilation $p_1=7\%$ when the stress is in between the low-and high-stress thresholds, and has a slope of less than -1000 Pa/s.  For all such events there is a time within the 3 seconds before the identified stress drop threshold where neither bright spot is dilated during that period.   Since the stress drops typically last less than 1 s -- comparable to the relative timing error -- we cannot calculate statistics for extended time periods of stress drop.  This indicates a correspondence between the stress drop and a lack of dilation, similar to the low-stress state.

\subsubsection{Size of the dilated region}

While we observe dilation at a single point on the surface in the high-stress state $p_1=52\%$ of the time, it is reasonable to ask if dilation at some point on the surface is required at all times to maintain the high-stress state.  Mechanically, it that would be required of a model in which the measured shear stress comes from friction transmitted through contact networks in the dilated region, in which the normal force is limited by surface tension at the liquid-air interface \cite{BJ12}.

 We infer a typical size of the dilated region by comparing the probability $p_1$ of dilation at a single point  with the probability  $p_2$ of dilation simultaneously at both observation points on the surface.    Probabilities of $p_1$ and $p_2$ are shown in Table \ref{tab:dilation_prob} for the different stress conditions.  $p_2$ is even more distinct between different stress conditions shown in the table than $p_1$,  suggesting the size of dilated regions may be a significant factor in determining stress states. 
 
 If we assume the dilated regions have a distribution of sizes characterized by a correlation length $\lambda$, and dilation events appear in the observation region one at a time, the probability of finding dilation at different spots of size $L=5$ mm separated by a distance $s=7.5$ mm at the same time can be estimated as
\begin{equation}
\frac{p_2}{p_1} = \exp\left(-\frac{s+L}{\lambda+L}\right) \ .
\label{eqn:corrlength}
\end{equation}
 Rearranging Eq.~\ref{eqn:corrlength} yields the correlation length $\lambda$, which is shown in Table \ref{tab:dilation_prob} for difference stress conditions. In the stress growth state, the extrapolated correlation length of 240 mm is larger than the circumference of 157 mm, suggesting that the size of dilated region during the stress growth period is likely to cover much of if not the entire surface.  This is consistent with the conclusion that the stress growth period typically results in surface-spanning dilation.  On the other hand, the extrapolated correlation length in the high-stress state  is only 29\% of the circumference.    This is consistent with the observation that we do not always observe dilation during the high-stress state, and it suggests a single dilation event may only be covering an average of 29\% of the surface in the high-stress state.  The fact that we observe $p_1 =52\% > 29\%$ in the high-stress state suggests that there is an average of more than one dilation event at a time on the surface in the high-stress state, or that the distribution of dilation event sizes is not well-represented by a typical correlation length as expressed in Eq.~\ref{eqn:corrlength}.  While we cannot guarantee with our limited observation locations that there is always dilation at some point of the surface in the high-stress state, the two-point statistics are consistent with that hypothesis for a reasonable assumption of how dilation events at two measuring points are correlated.

\subsubsection{Data sets with no high-stress state}

Three of the packing fractions we took measurements at were in the CST range, although due to instabilities at the surface, we could not take long measurements above $\tau_{max}$.  Instead, we took measurements at shear rates within the shear thickening transition ($\dot\gamma_c < \dot\gamma <\dot\gamma_{max}$).  Since these measurements were at both lower packing fractions and lower relative shear rates than the measurements where the high-stress state was observed, we cannot draw systematic conclusions regarding the difference in dilation between CST and DST at this time, but we can still test the relation between dilation and stresses.  At $\phi_{wt}/\phi_c=0.84$ and shear rate $\dot\gamma=125$ s$^{-1}$, no dilation events were observed over 145 s, and the shear stress did not fluctuate much above its mean of 40 Pa (this is well into the DST stress range which starts at $\tau_c = 2\pm1$ Pa for all packing fractions).  In the other two datasets at $\phi_{wt}/\phi_c=0.85$, $\dot\gamma=75$ s$^{-1}$ and $\phi_{wt}/\phi_c=0.89$, $\dot\gamma=20$ s$^{-1}$, we observed 2078 surface deformation events where part of the bright spot dimmed with $p_1=4\%$, but no stress fluctuations much larger than their respective averages of 27 Pa and 28 Pa.    While there are numerous surface deformation events in these cases, they are brief, so the probability $p_1$ is small.   The probability of both bright spots dimming at the same time drops to $p_2=0.08\%$, corresponding to a nearly zero correlation length according to Eq.~\ref{eqn:corrlength}.  These probabilities suggest very localized surface deformation events -- indeed in many cases we observe dimmed regions to be smaller than the bright spots used for measurement regions.   Thus, while surface deformation events can occur in flows that do not reach the high-stress state, these events are too small or brief to produce the surface-spanning dilation required for stress growth to reach the high-stress state.

\subsection{Viscosity curves with dilation }
\label{sec:visccurve}

\begin{figure} 
\centering
\includegraphics[width=0.475\textwidth]{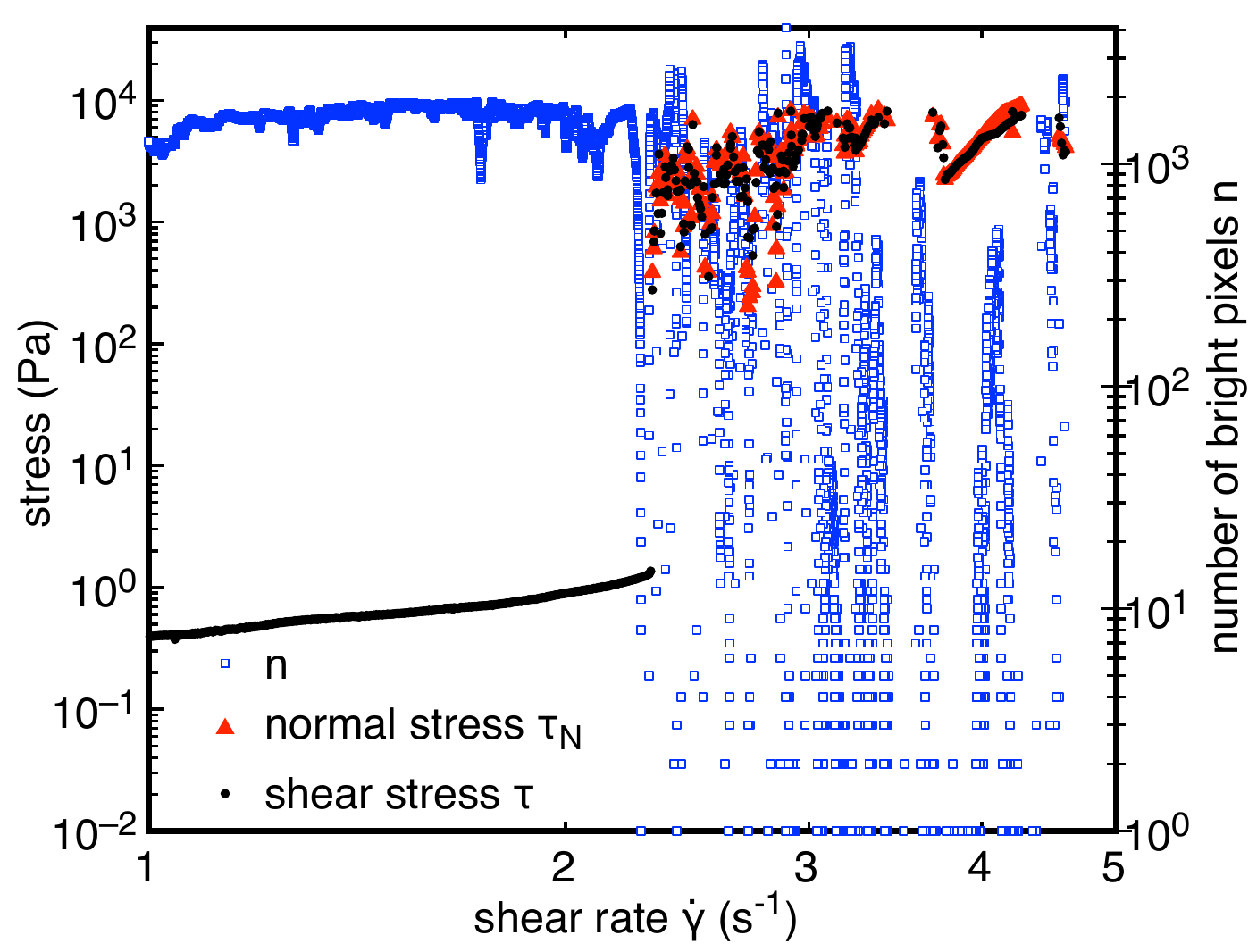}
\caption{A viscosity curve showing shear and normal stress (left axis)  and number of bright pixels $n$ on the surface (right axis), as a function of shear rate $\dot\gamma$.  The high  stress range coincides with the range of intermittent dilation events.  
 }
\label{fig:visccurve}
\end{figure}

The relation between dilation and shear thickening can be observed in a viscosity curve at $\phi_{wt}/\phi_c=0.94$ and suspension thickness $d=0.495$ mm.   Measurements of shear stress $\tau$, normal stress $\tau_N$, and the number of bright pixels $n$ are shown as a function of shear rate in Fig.~\ref{fig:visccurve}.  Stresses are reported every second, and number of bright pixels is reported 24 times per second so that fluctuations can be seen.  There are some gaps in the stress data where the shear rate left steady state because the stress reached the limit of what the rheometer could provide.    In the high-stress state at high shear rates, the normal and shear stress remain strongly coupled \cite{LDHH05}, and there are large fluctuations in the number of bright pixels to low values, indicating intermittent dilation. Note that the short timescale stress fluctuations are less visible than in Fig.~\ref{fig:timeseries} due to the longer averaging time of the stress data.   At shear rates below the shear thickening transition, there are no fluctuations in the number of bright pixels, and a complete lack of dilation is observed.  This confirms that the intermittent dilation occurs only in the high-stress state, and not at the low-stress state.
 
 We repeated this experiment 4 times with ramps of both decreasing and increasing shear rate, at 2 different packing fractions; in each case dilation was observed only in the high-stress state, and the transition between intermittent dilation and no dilation happened within a few seconds of crossing the shear thickening transition. 

\subsection{Dilation probability as a function of applied shear stress}
\label{sec:dilation_stress}

Since the shear thickening transition is sharp in shear rate, there is not much opportunity to observe how dilation evolves in the shear thickening range from that perspective.  On the other hand, the shear thickening transition exists over a non-zero range of stress in stress-controlled measurements.  In this subsection, we report statistics as a function of shear stress in stress-controlled measurements at a fixed packing fraction to observe the evolution of dilation through the shear thickening range.  

\begin{figure}
\centering
\includegraphics[width=0.475\textwidth]{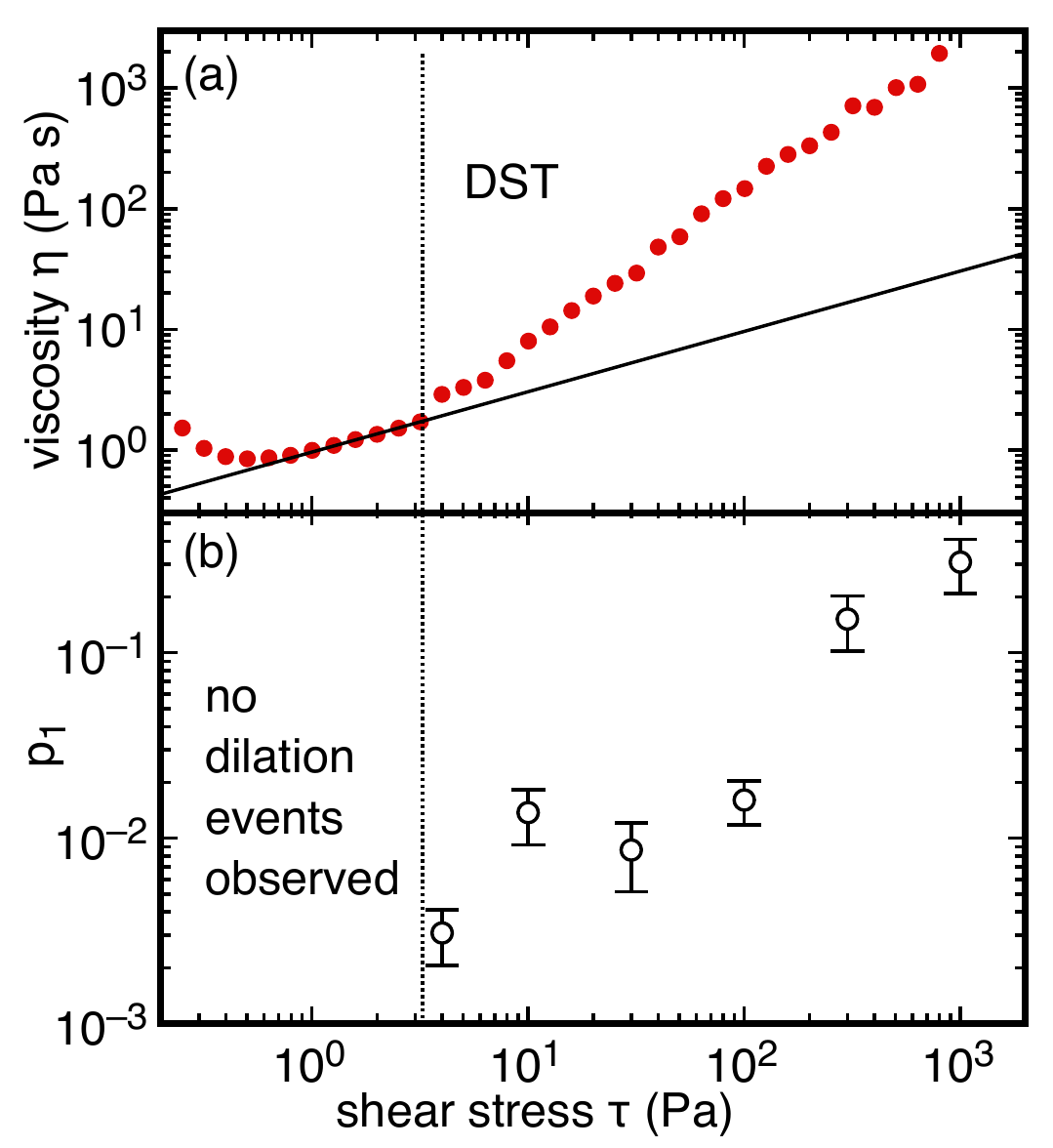}
\caption{(a) A viscosity curve $\eta(\tau)$.  Solid line: line of slope 1/2 used to determine the onset of DST.  Vertical line:  $\tau= \tau_c$, corresponding to the onset of DST.   (b) The probability of dilation $p_1$ as a function of controlled shear stress $\tau$.    Dilation is found only for $\tau > \tau_c$, and the probability of dilation increases with stress in the shear thickening range. 
}
\label{fig:probdilation_stress}
\end{figure}

To identify the stress range of shear thickening,  a  viscosity curve from stress-controlled measurements is shown in Fig.~\ref{fig:probdilation_stress}a for  $\phi_{eff}/\phi_c = 0.905$, at the low end of the DST range.  While $\eta(\dot\gamma)$ has a discontinuous jump at this packing fraction \cite{MB19}, DST in stress controlled measurements is characterized by  slopes of $\log\eta(\tau)$ approaching 1 in the limit as $\phi_{eff}$ approaches $\phi_c$ \cite{BJ09, BZFMBDJ11}.   For purposes of distinguishing DST in stress-controlled measurements from other shear thickening, we identify the onset of DST as the lowest stress value where the local value of $\partial\eta/\partial\tau$ exceeds 1/2 \cite{BJ12}.   DST is found for $\tau> \tau_c = 3$ Pa,  indicated by the vertical dashed line  in Fig.~\ref{fig:probdilation_stress}a.   
 
The probability of dilation $p_1$ is shown for constant applied stress measurements in Fig.~\ref{fig:probdilation_stress}b for similarly prepared suspensions as that shown in Fig.~\ref{fig:probdilation_stress}a (with a variation of $\phi_{eff}/\phi_c$ up to 0.008  based on measurements of $\dot\gamma_c$).   The plotted errors include the Poisson error assuming distinct (i.e.~non-continuous) dilation events are independent, plus a fractional error of 18\% due to the small variation in $\phi_{eff}/\phi_c$, which is calculated from an extrapolation of a comparison of the probability of dilation for the different time series reported in Table \ref{tab:dilation_prob}.  For $\tau = 1$ Pa ($<\tau_c$) where there is no shear thickening, we find zero dilation events.  For comparison of statistics, this measurement was made over the same time period that 44 dilation events were observed at $\tau=4$ Pa ($> \tau_c$).  For reference, the distributions $p(n/n_p)$ are shown in Fig.~\ref{fig:probnumpixels_stress} for $\tau=1$ Pa and $\tau=1000$ Pa.  For $\tau>\tau_c$, the probability of dilation typically increases with $\tau$, indicating another correlation between dilation and higher stress in the shear thickening range.   The increasing probability of dilation with increasing stress in the shear thickening range may correspond to an increasing correlation length for surface dilation with increasing stress in the shear thickening range, as was found in Table \ref{tab:dilation_prob}, and for particle velocities \cite{Heussinger13}.  These results confirm that dilation occurs in the same stress range as DST (where $\tau>\tau_c$), but dilation is not found at lower $\tau$.

\section{Discussion}
\label{sec:discussion}

\subsection{Stress growth and failure}

The repeated growth of stress to a plateau with fluctuations seen in Fig.~\ref{fig:timeseries} is reminiscent of stress-strain curves of other disordered granular materials and soils that dilate.  In such systems, the plateau represents the ultimate strength of the material, and fluctuations around a plateau are characteristic of a failure regime, where drops in stress correspond to localized failures.  Such localized failures have been observed as system-spanning force networks that repeatedly form and break up in the steady state in simulations of DST suspensions \cite{SMMD13, MSMD14}.  The observation that dilation covers a significant fraction of the surface  in this high-stress state is consistent with a picture in which the dilated surface provides the normal force to transmit stress along these localized  structures.  Strains of order 1 are typically expected for such failures in granular materials or dense suspensions \cite{FBOB09}, and the stress growth rate observed in Fig.~\ref{fig:timeseries} and our other datasets indicates that it typically takes a strain of $3.5\pm0.5$ to reach the failure stress plateau, on the expected order.  In contrast, during stress growth up to this plateau stress, we infer from Table \ref{tab:dilation_prob} that dilation tends to cover the entire surface by the end of the stress growth regime, more than in the high stress plateau.  Since the  stress is still growing, this suggests the force network has not yet reached a point of failure and remains connected, and does not start to fail locally until the stress plateau is reached.   

The drop in stress from the high-stress state down to the low-stress  state that we observe in Fig.~\ref{fig:timeseries} is a feature that is not typically observed in stress-strain curves of disordered granular materials and soils.   This indicates that the structure formed in the high-stress of DST state fails completely to support any stress on occasion and reverts to a fluid-like state.  It is possible for this melting to occur, and for the particles to rearrange and start the process of stress growth over again because this is a closed system.  However, the possibility and even likelihood that the structures fail completely indicates that the structures formed in DST suspensions are much more fragile than a dry granular material or soil, which do not suffer such complete failures in a similar shear. 


\subsection{Reiteration of the mechanism that connects surface dilation and stress in DST}

The observation of deformations at the surface of suspensions during DST, particularly when the stresses are high or growing, confirms the  coupling between stress and dilation in the high-stress state of DST.  The mechanism can be explained as follows \cite{BJ12}.  As the particle packing dilates in response to shear for $\tau>\tau_c$, it pushes against the boundaries of the suspension.  Since the liquid-air interface is the softest boundary, it determines the overall response.  As dilation deforms the liquid-air interface, surface tension provides a restoring force that pushes back on the particles with a maximum stress on the order of $\tau_{max}$ that scales with surface tension over particle size \cite{CHH05}.  This restoring force is transmitted along an effectively frictional contact network where shear stress is proportional to normal stress, which provides the resistance to shear in the high-stress state that is observed in a viscosity curve as the maximum stress $\tau_{max}$ of the shear thickening range.

While the connection between dilation and the high-stress state of DST has been shown here only for cornstarch and water, similar conclusions are likely true for suspensions that exhibit similar DST, specifically those which are found to exhibit positive normal stresses, visible surface dilation (although it is not yet clear if this is visible by eye for smaller particles as they become comparable to the wavelength of light), and where the maximum stress in the shear thickening range $\tau_{max}$ scales with surface tension divided by particle size.  Notably, the scaling of $\tau_{max}$ was confirmed for dozens of suspensions \cite{BJ12}.


\subsection{Bulk vs.~boundary rheology}

The fact that dilation at the boundary determines stresses means that the rheology of DST suspensions depends on the boundary conditions, rather than being a bulk material property.  When the suspension dilates to push against the boundary, it is the softest part of the system that determines the stiffness of the whole.   Usually, the liquid-air interface is the softest boundary in a rheometer or other open flows, which is why it determines $\tau_{max}$.  On the other hand, bulk simulations usually artificially remove boundaries by using periodic boundary conditions in fixed volumes, where instead the particle stiffness provides a restoring force against dilation instead of a boundary  \cite{Heussinger13, SMMD13, MSMD14, SGM17}.  Modified rheometer experiments have been demonstrated with solid boundaries showing $\tau_{max}$ scales with the boundary stiffness \cite{BJ12}.  It is for this reason that industrial flows of dense slurries and granular materials are often intentionally left with a free surface, rather than forced through pipes with hard walls which would produce a much larger resistance to shear.

The importance of boundary conditions in the rheology of DST suspensions also means that shear thickening is not an inherent feature of the constitutive relation of DST suspensions.  Modified rheometer experiments on the same suspensions have been demonstrated to show, for example, a flow with a constant normal stress instead of a constant suspension thickness results in a shear thinning rheology at all shear rates instead of shear thickening \cite{BJ12}.  

A  common observation in DST suspensions is the spilling of the suspension as the liquid-air interface becomes unstable at high shear rates.  This happens even when shear rates are low at high packing fractions where inertia is negligible.  This conditions for spillage are observed to be dependent on measurement procedure, but this tends to happen near $\tau_{max}$, which is why reported measurements of DST curves often end near $\tau_{max}$ and almost never go far beyond it.  This can be understood if the normal stress in DST suspensions for stable flows is limited by surface tension at the liquid-air interface:  a larger normal stress induced by shear resulting in dilation larger than about 1 particle diameter results in particles breaking through the surface, so the sample spills.

The Wyart-Cates (WC) model for shear thickening uses a different description of the high-stress state of a DST suspension, which is a bulk hybrid viscous-frictional state where shear stress is nearly proportional to both shear rate and and normal stress \cite{WC14}.    That model characterizes the high-stress state as originating from a bulk viscosity, rather than the  observed surface-dilation-dependent stress.   The high-stress state of the WC model differs from observations of many DST suspensions where the shear stress in the high-stress state is proportional to normal stress but largely independent of shear rate \cite{Ba89, BJ12}.  The bulk rheology of the standard WC model also fails to explain observations of different rheologies with different boundary conditions and spillage mentioned in the preceding paragraphs \cite{BJ12}.  It is likely that the WC model could be modified to include the  dilation-dependent mechanics of the high-stress state for these suspensions in place of the hybrid viscous-frictional state, and still keep the other important features of the model such as the transition between high and low-stress states, and a crossover function to describe viscosity curves.

\section{Summary}
Dilation can be observed on the surface of a suspension as a speckled pattern, as light reflects diffusively off a surface that is rough on the particle scale, which is due to dilation pushing particles through the liquid-air interface (Fig.~\ref{fig:raw_images}, Supplementary Videos 1, 2).  Dilation in steady state flows of DST suspensions is found to be intermittent and localized (Figs.~\ref{fig:dilation_stages}, \ref{fig:raw_pixels},  \ref{fig:timeseries}), and localized dilated regions can propagate along the surface of the suspension (Fig.~\ref{fig:frontpropagation}, Supplementary Video 2). 

 Dilation is strongly correlated to the shear and normal stress in the system, which fluctuate between high- and low-stress states (Fig.~\ref{fig:timeseries}).  Dilation is found 52\% of the time at a given observation point in the high-stress state of DST suspensions, and two-point statistics suggest dilation is likely to be found at some point on the surface in high-stress state at all times.  On the other hand,  a complete lack of dilation is observed in the low-stress state when there is not rapid stress increase within the following 9 s  (Table \ref{tab:dilation_prob}).   
During the stress growth from the low-stress state to the high-stress state, dilation is even more likely than in the high-stress state, and is observed 100\% of the time following 9 s of persistent stress growth (Table \ref{tab:dilation_prob}).   The 9 s periods seems to correspond to the maximum time required for signals relating dilation and stress to travel across the system.  The stress growth period can be interpreted similar to the growth portion of a material stress-strain curve before failure, after which dilation is still found, but more localized in the high-stress plateau where failure occurs intermittently.  While surface deformations are observed in some experiments where the high-stress state is not reached, the events are very localized, and so they do not produce the surface-spanning dilation required to reach that of the high-stress state (Table \ref{tab:dilation_prob}). 
 
We confirm that dilation occurs in both stress- and rate-controlled measurements in the stress range of shear thickening and the high-stress state of DST suspensions ($\tau>\tau_c$), but not in the low-stress state (Fig.~\ref{fig:visccurve}, \ref{fig:probdilation_stress}).  The numerous relationships found between dilation and the high-stress stress state indicate that dilation at the suspension-air interface is coupled to the stresses in the high-stress state of DST.  This coupling is explained by the argument that dilation causes particles to poke through the liquid-air interface, which responds with a restoring force from surface tension that is transmitted along a frictional particle contact network to provide the resistance to shear in the high-stress state \cite{BJ12}

 \section{Acknowledgements}

This work was supported by the NSF through grant DMR 1410157.   R. Maharjan carried out initial experiments, identified the algorithm for tracking dilation, and collected data for Figs.~\ref{fig:raw_images}--\ref{fig:probnumpixels_stress}, \ref{fig:probdilation_stress}.  E. O'Reilly and T. Postiglione carried out experiments with 2 cameras (Fig.~\ref{fig:timeseries} and Table \ref{tab:dilation_prob}).  N. Klimenko measured viscosity curves with video (Fig.~\ref{fig:visccurve}).  R. Maharjan, E. O'Reilly, T. Postiglione, and N. Klimenko carried out the bright spot tracking and manually checked the algorithm on their respective data.  E. Brown conceived the project, performed statistical analyses, and wrote the paper.

 \section{Supplementary Videos}
 
 Supplmentary videos can be found at:\\
 
 https://www.eng.yale.edu/brown/publications.html


%

\end{document}